\documentclass[floatfix,aps,tightenlines,prd,nofootinbib,superscriptaddress,onecolumn,12pt]{revtex4-1}

\usepackage{graphicx}
\usepackage{bm}
\usepackage{amsmath}
\usepackage{mathtools}

\newcommand{\W}{\ensuremath{\mathbf{W}}}

\newcommand{\Wmn}{\ensuremath{\mathbf{W}}^{[n]}_{(m)}}

\newcommand{\f}[1]{\ensuremath{{f}_{#1 } } }
\newcommand{\ti}[1]{\ensuremath{\mathbf{F}_{#1 } } }
\newcommand{\T}[1]{\ensuremath{\mathbf{T}_{#1 } } }

\newcommand{\wkb}{\ensuremath{W_{\{k,b\}}}}

\begin{document}
\title{Generic construction of scale-invariantly coarse grained  memory}
\author{Karthik H. Shankar}
\affiliation{Center for Memory and Brain, Boston University}

\begin{abstract}
Encoding temporal information from the recent past as spatially distributed activations is essential in order for the entire recent past to be simultaneously accessible. Any biological or synthetic agent that relies on the past to predict/plan the future, would be endowed with such a spatially distributed temporal memory. Simplistically, we would expect that resource limitations would demand the memory system to store only the most useful information for future prediction.
For natural signals in real world which show scale free temporal fluctuations, the predictive information encoded in memory is maximal if the past information is scale invariantly coarse grained.
Here we examine the general mechanism to construct a scale invariantly coarse grained memory system.  Remarkably, the generic construction is equivalent to encoding the linear combinations of  Laplace transform of the past information and their approximated inverses. 
This reveals a fundamental construction constraint on memory networks that attempt to maximize predictive information storage relevant to the natural world.   
\end{abstract}

\maketitle{}

\section{Introduction}

Representing the information from the recent past as transient activity distributed over a network has been actively researched in biophysical as well as purely computational domains \cite{MaassEtal02,Jaeger01}.  It is understood  that recurrent connections in the network can keep the information from distant past alive so that it can be recovered from the current state. The memory capacity of these networks are generally measured in terms of the accuracy of recovery of the past information \cite{Jaeger01,WhiteSompolinsky04,HermansSchrauwen09}. Although the memory capacity strongly depends on the network's topology and sparsity \cite{GanguliEtal08,StraussEtal12,LegeMaas07,WallaceEtal13}, it can be significantly increased by exploiting any prior knowledge of the underlying structure of the encoded signal \cite{GangSomp12,CharlesEtal14}.

Our approach to encoding  memory  stems from a focus on its utility for future prediction, rather than on the accuracy of recovering the past. 
%To bolster this motivation, it is worthy to note that unrelated prior works indicate that the energy efficiency of a memory system is directly related to the predictive information it stores \cite{StillEtal12}, and that energy efficiency and predictive relevance should be considered as fundamental biophysical principles  guiding the brain \cite{Friston10, Bialek_book}.
In particular we are interested in encoding time varying signals from the natural world into memory so as to optimize future prediction. It is well known that most natural signals exhibit scale free long range correlations \cite{Mand82,VossClar75,WestShle90}. By exploiting this intrinsic structure underlying natural signals, prior work has shown that   the predictive information contained in a finite sized memory system can be maximized if the past is encoded in a scale-invariantly coarse grained fashion \cite{ShanHowa13}.
Each node in such a memory system would represent a coarse grained average around a specific past moment, and the time window of coarse graining linearly scales with the past timescale. 
Clearly the accuracy of information recovery in such a memory system degrades more for more distant past. In effect, the memory system sacrifices accuracy in order to represent information from very distant past, scaling exponentially with the network size \cite{ShanHowa13}. The predictive advantage of such a memory system comes from washing out non-predictive fluctuations from the distant past, whose accurate representation would have served very little in predicting the future.   
%In lay terms--it is not important to accurately remember whether an event occurred 101 seconds in the past or 110 seconds in the past, while it is important to remember whether the event occurred 1 second or 10 seconds in the past.
Arguably, in the natural world filled with scale-free time varying signals, animals would have evolved to adopt such a memory system  conducive for future predictions. This is indeed evident from animal and human behavioral studies that show that our memory for time involves scale invariant errors which linearly scale with the target timescale \cite{Gibb77,RakiEtal98}.

Our focus here is not to further emphasize the predictive advantage offered by a scale invariantly coarse grained memory system, rather we simply assume the utility of such a memory system and focus on the generic mechanism to construct it. One way to mechanistically construct such a memory system is  to gradually encode information over real time as a Laplace transform of the past and approximately invert it \cite{ShanHowa12}. The central result in this paper is that any mechanistic construction of such a memory system is simply equivalent to encoding linear combinations of Laplace transformed past and their approximate inverses. This result should lay strong constraints on the connectivity structure of memory networks exhibiting the scale invariance property.

We start with the basic requirement that different nodes in the memory system represents coarse grained averages about different past moments. Irrespective of the connectivity, the nodes can be linearly arranged to reflect their monotonic relationship to the past time.  Rather than considering a network with a finite set of nodes, for analysis benefit, we consider a continuum limit where the  information from the past time is smoothly projected on a spatial axis. The construction can later be discretized and implemented in a network with finite nodes to represent past information from timescales that exponentially scale with the network size. 

\section{Scale Invariant Coarse Graining } 

\begin{figure}
\begin{center}
\includegraphics[width=0.55\textwidth]{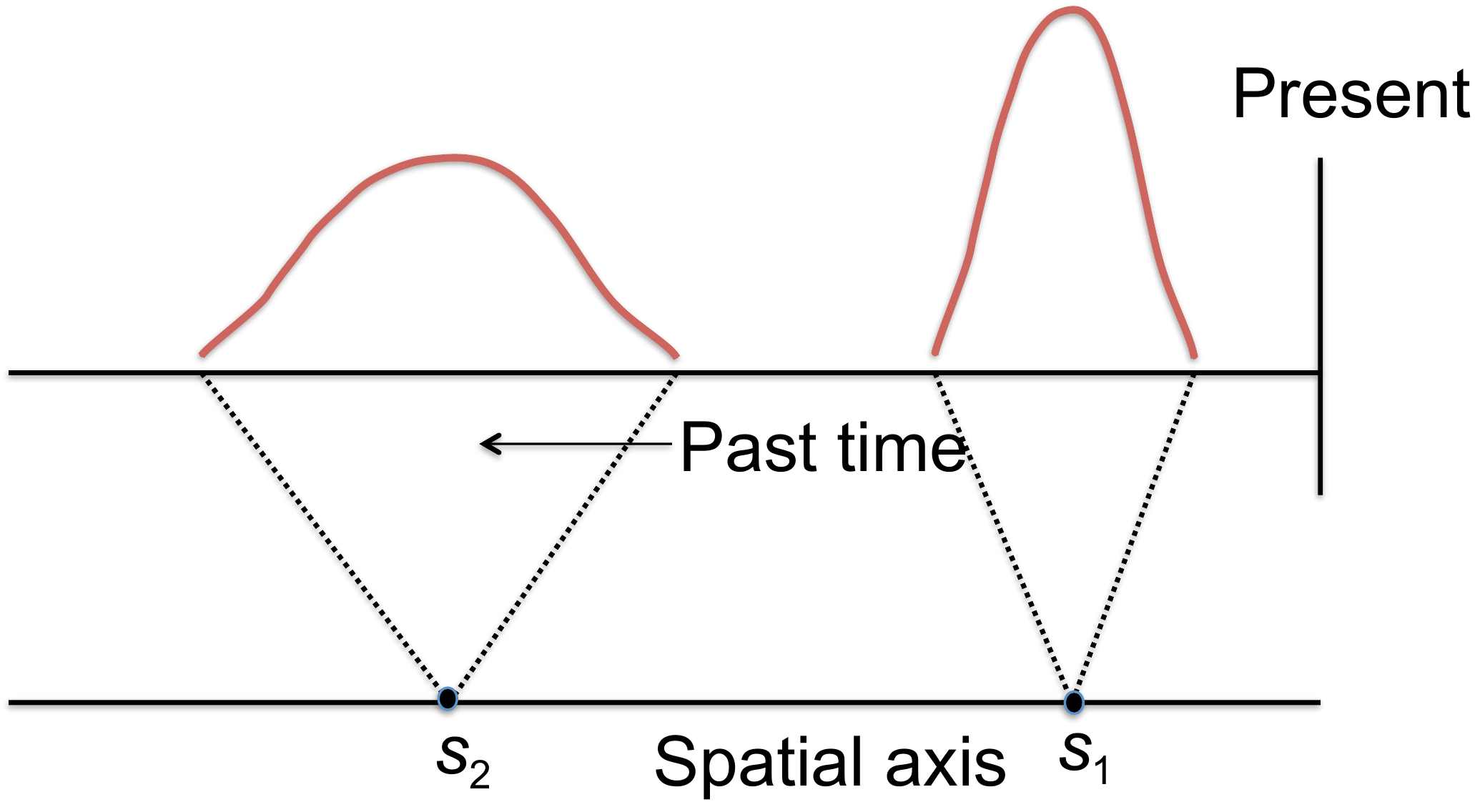}
\end{center}  
\caption{Coarse grained averages around different past instants are projected on to different points on the spatial axis.    
}
\label{cartoon}
\end{figure}

Consider a real valued function $\f{}(\tau)$ observed over time $\tau$. The aim is to encode this time-varying function into a spatially distributed representation in one dimension parametrized by $s$, such that at any moment $\tau$ the entire past from $-\infty$ to $\tau$ is represented in a coarse grained fashion as $\T{}(\tau,s)$
\begin{equation}
\T{}(\tau,s) = \int_{-\infty}^{\tau} \f{}(\tau') \W(\tau-\tau' ,s) \,\, d\tau'  .
\label{eq1}
\end{equation}
This is now a convolution memory model. The  kernel $\W{}(\tau-\tau',s)$ is the coarse graining window function  with normalized area for all $s$, $\int_{-\infty}^{\tau} \W(\tau-\tau',s) d\tau' =1$. Different points on the spatial axis uniquely and monotonically represents coarse grained averages about different instants in the past, as illustrated in figure~\ref{cartoon}.   

We require that coarse graining about any past instant linearly scales with the past timescale.  
So, for any pair of points $s_1$ and $s_2$, there exists a scaling constant $\alpha_{12}$ such that $\W(\tau-\tau',s_1)  = \alpha_{12} \W(\alpha_{12}(\tau-\tau'),s_2)$. For the window function to satisfy this scale-invariance property, there should exist a monotonic mapping $s(\alpha)$ from a scaling variable $\alpha$ to the spatial axis so that 
\begin{equation}
\W(\tau-\tau',s(\alpha)) = \alpha \W(\alpha (\tau-\tau'),s(1))  .
\end{equation}
Without loss of generality we shall pick $s(\alpha)=\alpha$ because it can be retransformed to any  other monotonic $s(\alpha)$ mapping after the analysis.  Hence with $0<s<\infty$,
\begin{equation}
\W(\tau-\tau',s) = s \W(s (\tau-\tau'),1)  .
\label{eqwindow}
\end{equation}

\section{Space-Time Local mechanism} 

Equation \ref{eq1} expresses the encoded memory as an integral over the entire past. However, the encoding mechanism can only have access to the instantaneous functional value of $\f{}$ and its derivatives. The spatial pattern should self sufficiently evolve in real time to encode  eq.~\ref{eq1}. This is a basic requirement to mechanistically construct $\T{}(\tau,s)$ in real time using any network architecture. Since the spatial axis is organized monotonically   to correspond to different past moments, only the local neighborhood of any point would affect its time evolution.  So we postulate that the most general encoding mechanism that can yield eq.~\ref{eq1} is a space-time local mechanism given by some differential equation for $\T{}(\tau,s)$. To analyze this, let us first express the general space-time derivative of $\T{}(\tau,s)$ by repeatedly differentiating   eq.~\ref{eq1} w.r.t $\tau$ and $s$. 
\begin{eqnarray}
\T{(m)}^{[n]} (\tau,s) &=& \sum_{j=0}^{n-1} f^{[n-j-1]} (\tau) \W^{[j]}_{(m)}(0,s) \nonumber \\
&+&  \int_{-\infty}^{\tau} \f{}(\tau') \W^{[n]}_{(m)} (\tau-\tau' ,s) \,\, d \tau'    .
\label{Tderiv}
\end{eqnarray}
Here $n$ and $m$ are positive integers. For brevity, we denote the order of time derivative within a square bracket in the superscript and  the order of space derivative within a parenthesis in the subscript.

Since $\f{}(\tau)$ is an arbitrary input, $\T{}(\tau,s)$ should satisfy a time-independent differential equation which can depend on instantaneous time derivatives of $\f{}(\tau)$. The first term in the r.h.s of eq.~\ref{Tderiv} is time-local, while the second term involves  an integral over the entire past. In order for the second term to be time-local, it must be expressible in terms of lower derivatives of $\T{}(\tau,s)$. Since the equation must hold for any $\f{}(\tau)$,  $\Wmn(\tau-\tau',s)$ should satisfy a linear equation. 

\begin{equation}
\sum_{n,m} C_{nm} (s) \Wmn (\tau-\tau',s) =0  .
\label{eq:comb}
\end{equation}   

The aim here is  not to derive the time-local differential equation satisfied by $\T{}(\tau,s)$, but just to impose its existence, which is achieved by imposing eq.~\ref{eq:comb} for some set of functions $C_{nm}(s)$. To impose this condition, let us first evaluate $\Wmn(\tau-\tau',s)$ by exploiting the functional form of the window function given by eq.~\ref{eqwindow}. Defining $z \equiv s(\tau-\tau')$ and the function $G(z) \equiv \W(z,1)$,  eq.~\ref{eqwindow} can be repeatedly differentiated to obtain  

\begin{equation*}
\Wmn(\tau-\tau',s) =   \sum_{r=r_o}^{m} (n+1)! m! \frac{s^{n+1-m+r}}{r! (m-r)!^2}  (\tau-\tau')^r G^{[n+r]}(z) ,
\end{equation*}
where $r_{o}=$ max[0, $m-n-1$] and the superscript on $G(z)$ represents the order of the derivative w.r.t $z$. Now eq.~\ref{eq:comb} takes the form
\begin{equation}
 \sum_{n,m}   C_{nm}(s) s^{n+1-m} \sum_{r=r_{o}}^{m}  \frac{(n+1)! m!}{r! (m-r)!^2} z^r G^{[n+r]}(z) =0
 \label{fin_cond}
\end{equation}

The above equation is not necessarily solvable for an arbitrary choice of  $C_{nm}(s)$. However, when it is  is solvable, the separability of the variables $s$ and $z$ implies  that  the above equation will be separable into a set of  linear differential equations for $G(z)$ with coefficients given by integer powers of $z$. The general solution for $G(z)$ is then given by
\begin{equation}
G(z)= \sum_{i,k} a_{ik} z^{k} e^{-b_i z},
\label{final_sol}
\end{equation}
where $i$ and $k$ are non negative integers. The coefficients $a_{ik}$ and $b_i$, and the functions $C_{nm}(s)$  cannot be independently chosen as they are constrained through eq.~\ref{fin_cond}. Once a set of $C_{nm}(s)$ is chosen consistently with the coefficients $a_{ik}$ and $b_i$, the differential equation satisfied by $\T{}(\tau,s)$ can be obtained by  iteratively substituting  $\Wmn(\tau-\tau',s)$ (in the second term of the r.h.s of eq.~\ref{Tderiv}) in terms of its lower derivatives and replacing the integral in terms of derivatives of $\T{}(\tau,s)$. 

Here we shall neither focus on the choice of $C_{nm}(s)$ nor on the differential equation for $\T{}(\tau,s)$ it yields. We shall only focus on the set of possible window functions that can be constructed by a space-time local mechanism. Hence it suffices to note from the above derivation that the general form of such a window function is given by  eq.~\ref{final_sol}.  Since by definition the window function at each $s$ coarse grains the input about  some past moment, we expect it to be non-oscillatory and hence restrict our focus to real values of $b_i$. Further, the requirement of the window function to have normalized area at all $s$ restricts $b_i$  to be positive.

\section{Two step process}

Let us consider the simplest window function, where only one of the coefficients in the set of $a_{ik}$ and $b_i$ in eq.~\ref{final_sol} are non-zero, namely  $b_i=b$ and $a_{ik} =b^{(k+1)}/k! $. We shall denote the corresponding window function as $\wkb$ to highlight its dependence on  specific  $k$ and  $b$. The most general window function is then simply a linear combination of various $\wkb$ for different values of $k$ and $b$. From eq.~\ref{final_sol}, $\wkb$ takes the form

\begin{equation}
 \wkb(\tau-\tau',s)  = \frac{(bs)^{k+1}}{k!} (\tau-\tau')^k e^{-bs(\tau-\tau')} .
 \label{wkb}
\end{equation}    
 %Note that  $\W^{[j]}_{(m)} (0,s) =0$ for any $k>j$. 
 It turns out that the differential equation satisfied by $\T{}(\tau,s)$ that generates this window function is  simply first order in both space and time given by  
\begin{equation}
\T{(1)}^{[1]} (\tau,s) + bs \T{(1)}^{[0]} (\tau,s) - \frac{(k+1)}{s} \T{(0)}^{[1]} (\tau,s) =0 ,
\label{diffeqT}
\end{equation}
with a boundary condition $\T{}(\tau,\infty) = \f{}(\tau)$. This equation can hence be evolved in real time by only relying on the instantaneous input $\f{}(\tau)$ at each moment $\tau$.     

%The only value $j$ can take in eq.~\ref{Tderiv} is zero because the maximum value of $n$ is 1. Hence the first term in the r.h.s of eq.~\ref{Tderiv} vanishes while the second term is expressed as lower derivatives of $\T{}(\tau,s)$, leading to a differential equation consisting only derivatives of $\T{}(\tau,s)$ and requiring the boundary condition $\T{}(\tau,\infty) = \f{}(\tau)$.  

For more complex window functions that are linear combinations of  $\wkb$ for various $k$ and $b$, the order of the space and time derivatives of $\T{}(\tau,s)$ involved in the differential equation are not necessarily bounded when the parameters $k$ and $b$ involved in the linear combinations of $\wkb$ are bounded.  So, it is not straight forward to derive the mechanistic construction as a differential equation for $\T{}(\tau,s)$. Hence the question now is, what is the mechanism to construct a memory system with any window function? 

Interestingly, there exists an alternative derivation of eq.~\ref{diffeqT} where the time derivative and space derivative can be sequentially employed in a two step process \cite{ShanHowa12}. The first step is equivalent to encoding the Laplace transform of the input $\f{}(\tau)$ as $\ti{}(\tau,s)$. The second step is equivalent to approximately inverting the Laplace transformed input to construct $\T{}(\tau,s)$. 

\begin{equation*}
\f{}(\tau)\,\, \underrightarrow{\,\,\,Laplace\,\,\, } \,\,  \ti{}(\tau,s)\,\, \underrightarrow{\,\,\,Inverse\,Laplace \,\,\,} \,\, \T{}(\tau,s)
\end{equation*} 
\begin{eqnarray}
\ti{}^{[1]}(\tau,s) &=& -bs\ti{}(\tau,s) + \f{}(\tau),   \label{smallt}  \\
 \T{}(\tau,s) &=& \frac{b}{k!} s^{k+1} \ti{(k)}(\tau,s) .  \label{bigT}
\end{eqnarray}

Taking $\f{}(\tau)$ to be a function of bounded variation and $\ti{}(-\infty,s)=0$, eq.~\ref{smallt} can be integrated  to see that $\ti{}(\tau,s)= \int_{-\infty}^{\tau} \f{}(\tau') e^{-bs(\tau-\tau')} d\tau' $. 
Thus $\ti{}(\tau,s)$ is the Laplace transform of the past input computed over real time. Eq.~\ref{bigT} is an approximation to inverse Laplace transform operation \cite{Post30}. So $\T{}(\tau,s)$ essentially attempts to reconstruct the past input, such that at any $s$, $\T{}(\tau,s) \simeq \f{}(\tau-k/bs)$. This reconstruction grows more accurate as $k \rightarrow \infty$, and the input from each past moment is reliably represented at specific spatial location. For finite $k$ however, the reconstruction is fuzzy and each spatial location represents a coarse grained average of inputs from past moments, as characterized by the window function $\wkb$. For further details, refer to \cite{ShanHowa12}.

Since any window function is a linear combination of various $\wkb$ for different values of $k$ and $b$, its construction is essentially equivalent to linear combinations of the two step process given by equations~\ref{smallt} and \ref{bigT}. 

\begin{figure}[htbp]
\begin{minipage}{0.45\textwidth}
  \begin{tabular}{|c | c | }
    \hline 
     & Combinations of $\wkb$  \\ \hline \hline
    1 & $W_{\{100,2 \}}$   \\
    \hline
    2 & $0.32W_{\{65,1.3\}} + 0.27W_{\{75,1.8\}} $ \\ 
        & $ +0.32W_{\{75,1.26\}}+0.09W_{\{100,1.62\}}  $  \\
    \hline
    3 & $ W_{\{ 8,0.16\}} $ \\
    \hline
     4 & $ 0.5W_{\{ 35,1\}} + 0.5  W_{\{70,1\}} $ \\
     \hline 
  \end{tabular}
 \end{minipage} 
 \begin{minipage}{0.5\textwidth}
\begin{center}
\includegraphics[width=1\textwidth]{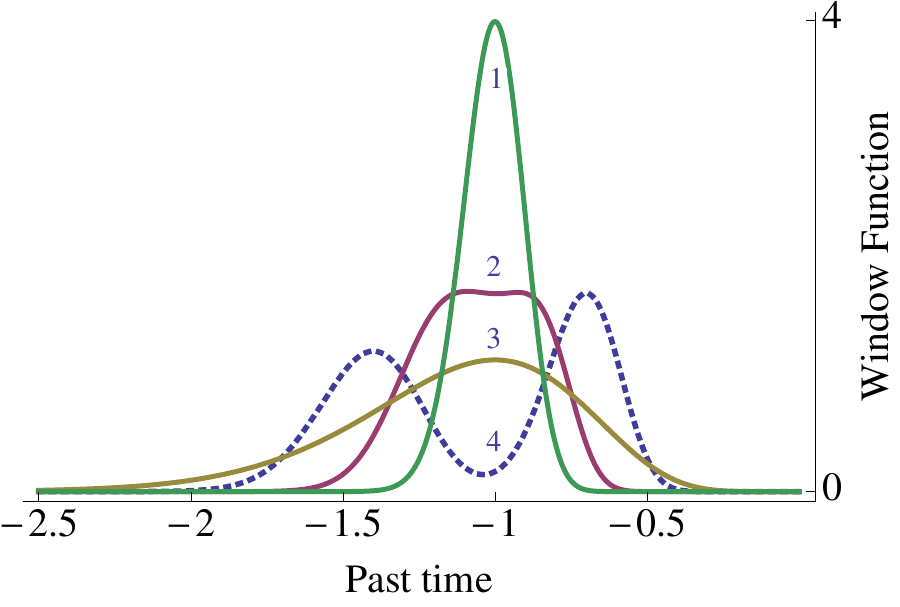}
\end{center}  
\end{minipage}
\caption{For different combinations of $\wkb$, the window functions are plotted as a function of past time at the spatial point $s=50$.}
\label{wind}
\end{figure}

The choice of the combinations of $\wkb$ has strong implications on the shape of the resulting window function.  At any given $s$, $\wkb$ is a unimodal function with a peak at $\tau-\tau' =k/bs$ (see eq.~\ref{wkb}). Arbitrary combinations of $\wkb$ could result in a spatial location representing the coarse grained average about disjoint past moments, leading to undesirable shapes of the window function. Hence the values of $k$ and $b$ should be carefully tuned. Figure \ref{wind} shows the window functions constructed from four combinations of $b$ and $k$.  The combinations are chosen such that at the point $s=50$, the window function coarse grains around a past time of $\tau' -\tau \simeq -1$. The scale invariance property guarantees that its shape remains identical at any other value of $s$ with a linear shift in the coarse graining timescale.  Comparing combinations 1 and 3, note that the window function is narrower for larger $k$(=100) than for a smaller $k$(=8).  Combination 2 has been chosen to illustrate a plateau shaped window function whose sides can be made arbitrarily vertical by fine tuning the combinations. Combination 4 (dotted curve in fig.~\ref{wind}) illustrates that combining different values of $k$ for the same $b$ will generally lead to a multimodal window function which would be an undesirable feature.   

\section{Discretized spatial axis}

A memory system represented on a continuous spatial axis is not practical, so the spatial axis should be discretized to finite points (nodes).  The two step process given by equations \ref{smallt} and \ref{bigT} is optimal for discretization particularly when the nodes are picked from a geometric progression in the values of $s$ \cite{ShanHowa13}.  Eq.~\ref{smallt} implies that the activity of each node evolves independently of the others to construct $\ti{}(\tau,s)$ with real time input $\f{}(\tau)$. This is achieved with each node recurrently connected on to itself with an appropriate decay constant of $bs$. Eq.~\ref{bigT}  involves taking the spatial derivative of order $k$ which can be approximated by the discretized  derivative requiring linear combinations of activities from $k$ neighbors on either sides of any node. For further details on implementation of the two step process on discretized spatial axis, refer to \cite{ShanHowa13}. 

 \begin{figure}[htbp]
\begin{center}
\includegraphics[width=.5\textwidth]{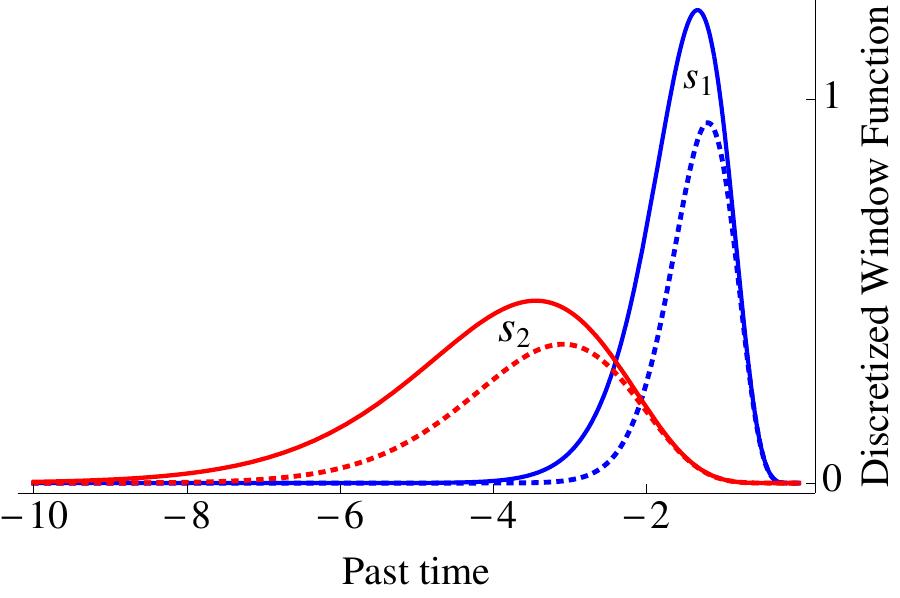}
\end{center}  
\caption{Window function $W_{\{ 8,1\}}$ at two points $s_1=6.72$ and $s_2=2.59$ computed on a discretized spatial axis with $c=0.1$. The dotted curves correspond to the window functions computed on the continuous spatial axis ($c \rightarrow 0$).}
\label{window_discrete}
\end{figure}

 By choosing the nodes along the $s$-axis  from a geometric progression, the error from the discretized spatial derivative  will be uniformly spread over all timescales, hence such a discretization is ideal to preserve scale-invariance.  Let us choose the $s$-values of successive nodes to have a ratio $(1+c)$, where $c < 1$. Figure \ref{window_discrete} shows the window function $\wkb$ with $k=8$ and $b=1$ constructed from the discretized axis with $c=0.1$. The window functions  at two spatial points $s_1=6.72$ and $s_2=2.59$ are plotted to illustrate that scale invariance is preserved after discretization. As a comparison, the dotted curves are plotted to show the corresponding window function constructed in the continuous $s$-axis (limit $c \rightarrow 0$). The window function computed on the discretized axis is artificially scaled up  so that the solid and dotted curves in figure \ref{window_discrete} are visually discernible.
 Note that the discretized window function peaks farther in the past time and is wider than the window function on the continuous spatial axis. As $c \rightarrow 0$, the discretized window function converges on to the  window function constructed on the continuous axis, while for larger values of $c$ the discrepancy grows larger. Nevertheless, for any value of $c$, the discretized window function always stays scale-invariant, as can be seen by visually comparing the shapes of the window functions at $s_1$ and $s_2$ in figure~\ref{window_discrete}.  Now, it is straight forward to construct scale-invariant window functions of different shapes by taking  linear combinations of discretized  $\wkb$, analogous to the construction in figure \ref{wind}.
 
 Implementing this construction on a discretized spatial axis as a neural network has a tremendous resource conserving advantage. Since at each $s$, the window function $\wkb$ coarse grains the input around a past time of $k/bs$, the maximum past timesscale represented by the memory system is inversely related to minimum value of $s$. The geometric distribution of the $s$ values on the discretized axis implies that if there are $N$ nodes spanning the spatial axis for $\T{}(\tau,s)$, it can represent the coarse grained past from timescales proportional to $(1+c)^N$. Hence exponentially distant past can be represented in a coarse grained fashion with linearly increasing resources.

\section{Discussion and Conclusion}

The formulation presented here starts from a convolution memory model (eq.~\ref{eq1}) and derives the form of the scale-invariant window functions (or the kernels) that can be constructed from a space-time local mechanism. Interestingly, by simply postulating a kernel of the form of eq.~\ref{final_sol}, Tank and Hopfield have demonstrated the utility of such a memory system in temporal pattern classification \cite{TankHopf87}.  
In general, a convolution memory model can adopt an arbitrary kernel, but it cannot be mechanistically constructed from  a space-time local differential equation, which means a neural network implementation need not exist. However, the Gamma-memory model \cite{Gamma92} shows that linear combinations of Gamma kernels, functionally similar to eq.~\ref{final_sol}, can indeed be mechanistically constructed from a set of differential equations. 

The construction presented here takes a complementary  approach to the Gamma-memory model by requiring scale invariance of the window function in the forefront and then imposing a space-time local differential equation to derive it. This sheds light on the connectivity between neighboring spatial units of the network that is required to generate a scale invariant window function, as described by the second part of the two step process (eq.~\ref{bigT}).  
 Moreover, the linearity of the two step process and its equivalence to the Laplace and Inverse Laplace transforms makes the memory representation analytically tractable.
 
Theoretically, the utility of a scale invariantly coarse grained memory  hinges on the existence of scale free temporal fluctuations in the signals being encoded \cite{ShanHowa13}. Although detailed empirical analysis of natural signals is needed to confirm this utility,  preliminary analysis of time series from sunspots and global temperature show that such a memory system indeed has a higher predictive power than a conventional shift register \cite{ShanHowa13}.  The predictive advantage of this memory system can be understood as arising from its intrinsic ability to wash out non-predictive stochastic fluctuations in the input signal from distant past and just represent the predictively relevant information in a condensed form. Finally, the most noteworthy feature is that a memory system with $N$ nodes can represent information from exponentially past times proportional to  $(1+c)^N$. In comparison to a shift register with $N$ nodes which can accurately represent a maximum past time scale proportional to $N$, this memory system is exponentially resource conserving.

%Hence, in the context of constructing memory networks with random recurrent connectivities, imposition of  scale-invariance will place significant constraints on their construction due to their equivalence to linear combinations of the described two step process.      
 
% Alternatively, we could envision storing the past information in an accurate way using a shift register and applying any window function on the stored past. However, with $N$ nodes in the shift register, only a maximum past timescale proportional to $N$ can be stored.  The accuracy of recovering past information from a shift register sharply drops to zero for timescales  beyond its capacity, while in recurrent and feedforward networks it smoothly decays to zero \cite{WhiteSompolinsky04,GanguliEtal08}. The two step process discussed here involves both recurrent and feedforward connectivities with tailored connection strengths. In the context of constructing memory networks with random recurrent connectivities, imposition of  scale-invariance will place significant constraints on their construction due to their equivalence to linear combinations of the described two step process.      

 %Hence, to the extent scale invariantly coarse graining the past has computational utility for memory, linear combinations of the two step process involving Laplace transform and its inversion is both generic and exponentially resource conserving.

\section*{Acknowldegements :} The work was partly funded by NSF BCS-1058937 and AFOSR FA9550-12-1-0369.

%\bibliography{newbib}


\end{document}